\def\be{\begin{equation}}
\def\ee{\end{equation}}
\def\smu{\sigma_{\mu}}

\def\Tr{\hbox{Tr}}
\def\sx{\sigma_x}
\def\sy{\sigma_y}
\def\sz{\sigma_z}
\def\epsaa{\varepsilon_{a\alpha}}
\def\epsbb{\varepsilon_{b\beta}}
\def\epsba{\varepsilon_{b\alpha}}
\def\epsinaa{\varepsilon_{\alpha a}}
\def\epsinbb{\varepsilon_{\beta b}}
\def\epsinba{\varepsilon_{\beta a}}

\def\CI{{\cal I}}
\def\OAa{O^A_a}
\def\OBb{O^B_b}
\def\OAb{O^A_b}
\def\PAa{P^A_{\alpha}}
\def\PAb{P^A_{\beta}}
\def\PBb{P^B_{\beta}}

\def\irt{\frac{1}{\sqrt{2}}}
\def\ket#1{| #1 \rangle}
\def\kett#1#2{| #1 , #2 \rangle}
\def\kettp#1#2#3{| #1, #2; #3 \rangle}
\def\kettt#1#2#3{| #1, #2, #3  \rangle}
\def\ketttp#1#2#3#4{| #1, #2, #3; #4 \rangle}
\def\braketbb#1#2#3#4{\langle #1,#2|#3,#4\rangle}
\def\ketAa{|A, \alpha \rangle}
\def\ketBb{|B, \beta \rangle}
\def\braAa{\langle A, \alpha|}
\def\naz{n^1_z}
\def\nbz{n^2_z}

\def\nbx{n^2_x}
\def\ncx{n^3_x} 
\def\nay{n^1_y}
\def\nby{n^2_y}

\def\nnaz{\bar{n}^1_z}
\def\nnbz{\bar{n}^2_z}

\def\nnbx{\bar{n}^2_x}
\def\nncx{\bar{n}^3_x}
\def\nnay{\bar{n}^1_y}
\def\nnby{\bar{n}^2_y}

\documentstyle[aps, prl, graphics, twocolumn, epsfig, floats]{revtex}

\title{Mutually unbiased binary observable sets 
on $N$ qubits}
\author{Jay Lawrence$^1$, \v Caslav Brukner$^2$, and Anton
Zeilinger$^2$ \\ {\it $^1$Department of Physics, Dartmouth
College, Hanover, NH 03755, USA,\\ $^2$Institute for
Experimentalphysics, University
of Vienna, Boltzmanngasse 5, A--1090 Vienna, Austria}}

\date{\today}

\begin{document}
\maketitle
\begin{abstract}
The Pauli operators (tensor products of Pauli matrices)
provide a complete basis of operators on the Hilbert space 
of $N$~qubits.   We prove that the set of $4^N-1$~Pauli 
operators may be partitioned into $2^N+1$~distinct subsets, 
each consisting of $2^N-1$ internally commuting observables.  
Furthermore, each such partitioning defines a unique choice 
of $2^N+1$~mutually {\it unbiased} basis sets in the 
$N$-qubit Hilbert space.   
Examples for 2 and 3 qubit systems are discussed with 
emphasis on the nature and amount of entanglement that 
occurs within these basis sets.  
\end{abstract}
\medskip
\centerline{PACS numbers:  03.67.-a, 03.65.Ta, 03.65.Ud}

\section{Introduction}
A pure quantum state of an $N$-qubit system is specified 
by the eigenvalues of $N$ independent commuting binary 
observables (``$N$ qubits carry $N$ bits of information'' 
\cite{az,bz1}).  In fact, a complete set of $2^N$ such 
states are so specified, each being associated with a
binary number consisting of the $N$ eigenvalues.  Of the
many alternative choices of observable sets that define
$N$-qubit basis sets, we are interested in those
that are maximally incompatible in the sense that a state
producing precise measurement results in one set produces
maximally random results in the other.

The Pauli operators \cite{nielsen} provide an explicit 
realization of points raised above.  First, they illustrate 
that although a greater number $2^N-1$ of observables 
simultaneously take definite values, only $N$ of these are
required to define a pure state, and in fact these $N$ 
generate all of the remaining compatible observables through 
multiplication.  On the other hand, all $4^N - 1$ Pauli 
operators are required in order to determine an arbitrary
mixed state.  In this connection, we shall show that the 
full set of these operators is exhausted in forming $2^N + 1$ 
distinct subsets, each consisting of $2^N-1$ internally 
commuting observables, and each defining its own unique
eigenbasis.  Both the observable sets and the corresponding
basis sets are called mutually unbiased \cite{wootters1} 
(and in previous works the observable sets have also been 
called mutually complementary \cite{bz1}) because of the 
following physical property:  If an $N$ qubit system is 
prepared in a joint eigenstate of 
one such observable set, then it has a uniform probability 
distribution over the joint eigenstates of any of the other 
sets.   It follows that all $2^N(2^N-1)$ observables outside
the original (maximal) set of $2^N-1$ compatible observables
will produce measurement results that are uniformly 
distributed over all possibilities.  Since the Pauli 
operators have binary spectra, it also follows that their 
dispersion is maximized.  

The equivalence of unbiasedness of basis sets and operator
sets may be understood from the formal definition as applied
to basis sets, which may be summarized in general terms as 
follows:  Let us denote basis sets by 
$A = 1,2,...$, and states within a basis by $\ketAa$, with 
$\alpha = 1,2,...,d$ (for the moment, we consider a Hilbert
space with general dimension $d$, although our interest here 
is in $d = 2^N$).  Two bases $A$ and $B$ are said to be 
mutually unbiased \cite{wootters1,ivanovic} if a system 
prepared in any element of $A$ (such as $\ketAa$) has a uniform 
probability distribution of being found in any element of $B$ 
\be
\label{unbiased1}
   |\braketbb{A}{\alpha}{B}{\beta}|^2 = d^{-1}, 
       \hskip0.7truecm (A \neq B),
\ee
where individual bases are understood to be orthonormal,
\be
\label{unbiased2}
 \braketbb{A}{\alpha}{A}{\beta} = \delta_{\alpha \beta}.
\ee
Certainty of measurement outcomes for the operator set 
defining the $\ketAa$'s implies a uniform probability 
distribution over states $\ketBb$, and this in turn implies
a uniform probability distribution over all eigenvalue sets
(and distinct measurement outcomes) of operators defining
the $\ketBb$'s.

A particular motivation for considering unbiased basis sets
is that they provide for the most efficient determination, 
using measurements alone, of
a general (pure or impure) quantum state \cite{wf}.   
In a $d$-dimensional Hilbert space, one needs $d^2-1$ real 
parameters to specify a general density matrix $\rho$, which 
must be hermitean and have $Tr(\rho) = 1$.  Since measurements
within a particular basis set can yield only $d-1$ independent
probabilities, one needs $d+1$ distinct basis sets to provide 
the required total number of $d^2 -1$ independent 
probabilities.  Ivanovi\'{c} \cite{ivanovic} showed that the 
required number $d+1$ of unbiased basis sets indeed 
exists if $d$ is a prime number, and Wootters and Fields 
\cite{wf} showed that it exists if $d$ is any {\it power} 
of a prime number.  Our proof is based upon this theorem of 
Wootters and Fields. 

The question of the existence and construction of unbiased
basis sets is interesting not only from a fundamental point 
of view (e.g., in the formulation of ``quantum mechanics 
without probability amplitudes'' \cite{wootters1}, and in 
the information-theoretic formulation of quantum mechanics 
\cite{bz1}), but also as an important ingredient in 
quantum-information protocols (e.g., in the solution of ``the 
mean king's problem'' \cite{aharonov} and in quantum 
cryptography \cite{crypto}).  In particular, it was found 
recently that key distributions based on higher-dimensional 
quantum systems with larger numbers of unbiased basis sets
can have certain advantages over those based on qubits 
\cite{pasq}.

The present paper illustrates how the study of operator 
relationships can provide a useful approach to the 
construction of unbiased basis sets of entangled as well as
product character.  The $N$-qubit Hilbert 
space has dimension $d = 2^N$, and operators on this space 
(which include the density matrix) live in their own vector 
space of dimension $4^N$.  The complete basis consisting of 
the Pauli operators \cite{nielsen} may be written as 
follows: Starting with the usual $2 \times 2$~Pauli matrices 
and the identity $I$ that act on the spaces of individual 
qubits,
\be
  \smu = (\sx,~\sy,~\sz,~I), \hskip0.7truecm  
   \mu = (1,~2,~3,~4),
\ee
we write the $4^N$ tensor products (the Pauli operators and 
identity $\CI$) that act on the $N$-qubit Hilbert space as 
\be
\label{pauli}
  O_i = \sigma^1_{\mu(1,i)} \sigma^2_{\mu(2,i)}...
  \sigma^N_{\mu(N,i)} =  \prod_{k=1}^N  \sigma^k_{\mu(k,i)}, 
\ee
where $k$ is the particle label and $i$ distinguishes among
the $4^N$ choices of the $N$ subscripts $\mu(k,i)$. This 
basis is $orthonormal$ \cite{orthonormal}; the inner product 
of two operators is defined as the trace of their product,  
\begin{eqnarray}
 \Tr (O_iO_j) &=& \prod_{k=1}^N \Tr
\Big( \sigma^k_{\mu(k,i)}
  \sigma^k_{\mu(k,j)} \Big) \nonumber \\ &=&
\prod_{k=1}^N  2 \delta_{\mu(k,i)\mu(k,j)} = 2^N \delta_{ij}, 
\label{orthonorm}
\end{eqnarray}
where $i=j$ means that $\mu(k,i) = \mu(k,j)$ for every 
particle $k$.  Like the individual Pauli matrices, each 
tensor product is self-inverse, $O_i^2 = \CI$, and apart 
from the identity (for which we reserve $i = 4^N$ so that 
$O_{4^N} = I^1 I^2...I^N \equiv \CI$) they are all traceless 
and have eigenvalues $\pm 1$.  

The binary spectrum for each observable $O_i$ permits its
expression as a binary proposition:  The 
two eigenvalues $\pm 1$ of the observable $O_i$ correspond
to the values ``true'' or ``false'' of the proposition  
``The product of the spin projections $\sigma^1_{\mu(1,i)} 
\sigma^2_{\mu(2,i)}...\sigma^N_{\mu(N,i)}$ is +1.'' (If a
particular $\sigma^k_{\mu}$~happens to be the identity, then
no statement is made about the $k$th qubit.)  

\section{General results for $N$ qubits}

We may now proceed to demonstrate the main formal points of
the paper:  First, that the set of $4^N -1$~Pauli operators 
(excluding the identity) may be partitioned into
$2^N +1$~subsets, each consisting of $2^N -1$~internally 
commuting members, and second, that every such partitioning 
defines a unique choice of unbiased basis sets (i.e., there 
is a one-to-one mapping from partitionings to choices of 
unbiased basis sets).

The first part makes use of the proven existence of $2^N +1$ 
unbiased basis sets \cite{wf}.  The projectors onto the 
unbiased basis states, 
\be
\label{projection1}
   \PAa = \ketAa \braAa,
\ee
may be used to re-express the unbiasedness of bases 
$A \neq B$ (Eq. \ref{unbiased1}) as
\be
\label{projection2}
   \Tr(\PAa \PBb) = 2^{-N}.
\ee
and to define a set of operators 
$\OAa$~by their spectral decompositions
\be
  \OAa = \sum_{\alpha = 1}^{2^N} \epsaa \PAa.
\label{spectral}
\ee
We {\it define} $\epsaa$ as a $2^N \times 2^N$ matrix consisting 
of orthogonal row vectors, one of whose entries are all +1's, 
and the remaining of whose entries are equal numbers of $+ 1$'s 
and $- 1$'s. There are exactly $2^N$ such orthogonal vectors, the 
components of each vector $a$ being the eigenvalues of $\OAa$.
One of these operators (say the $a = 2^N$th) is proportional to 
the identity, $O^A_{2^N} = \CI$.  We include this to make Eq. (8) 
invertible, which will be useful later.  The columns $\alpha$
label the joint eigenstates of the $\OAa$ ($a = 1,2,...,2^N-1$),
and comprise the truth tables associated with the $2^N-1$
corresponding propositions.  This labeling is redundant; clearly 
an appropriate subset of just $N$ rows may be used to construct 
$N$-component column vectors that define all $2^N$ joint
eigenstates unambiguously as binary numbers.  This reflects a
property of the Pauli operators mentioned earlier. 

The above definition provides $2^N+1$ distinct sets (indexed by 
$A$), each set containing $2^N-1$ operators (fixed $A$ and 
running index $a = 1,...,2^N-1$), after discarding the identity.  
Each of these operators has the spectrum $\pm 1$ and is 
traceless, by construction.  To show that they are unitarily 
equivalent to the Pauli operators, we need only demonstrate 
that they form an orthonormal set.  For the case $A \neq B$,
\be
   \Tr(\OAa\OBb) =  \sum_{\alpha,\beta} \epsaa \epsbb     
   \Tr(\PAa \PBb) = 0,
\ee
where Eq. (\ref{projection2}) and the property 
$\sum_{\alpha} \epsaa = 0$ were used; 
then for the case $A = B$, using Eqs. (\ref{projection1})
and (\ref{unbiased2}),
\begin{eqnarray}
   \Tr(\OAa\OAb) &=&  \sum_{\alpha,\beta} \epsaa \epsbb
   \Tr(\PAa \PAb)  \nonumber \\
     &=& \sum_{\alpha} \epsaa \epsba = 2^N  \delta_{ab}.
\end{eqnarray}
Finally, this orthonormal set of $4^N-1$ traceless operators 
is completed by adding the identity, so indeed they have a 
representation in the form given by Eq. (\ref{pauli}).
This shows that the Pauli operators may be partitioned 
accordingly.

We now show the second part, namely, that {\it any} such 
partitioning of Pauli operators defines a unique choice of 
unbiased basis sets.  Assuming such a partitioning, each subset 
($A$) of Pauli operators $\{O^A_1,O^A_2,...,O^A_{2^N-1}\}$ 
defines a unique basis of $2^N$ joint eigenstates $\ketAa$, 
$\alpha =1,...2^N$.  Thus, each $\OAa$ operator may be expanded 
as in Eq. (\ref{spectral}), with $\epsaa$ now defined as the 
eigenvalue of $\OAa$ on the state $\ketAa$, the lower index 
taking the values $a = 1,...,2^N-1$.  The known spectrum of the
$\OAa$'s dictates that each row of the $\epsaa$ matrix must 
consist of an equal number of $+1$'s and $-1$'s, and the 
identity Tr$(\OAa \OAb) = 2^N \delta_{ab}$ shows (note Eq. 10) 
that any two rows $a$ and $b$ are orthogonal.  Thus, by 
appending an additional row ($a=2^N$) to the $\epsaa$ matrix, 
we recover its previous form.  The scaled matrix 
$\epsaa/\sqrt{2^N}$ is orthogonal, and, therefore, we may 
invert Eq. (\ref{spectral}) to yield the projection operators 
\be
\label{Pexpand}
   \PAa = 2^{-N} \sum_a \epsinaa \OAa = 2^{-N} \Big(\CI +
   \sum_a^{'} \epsinaa \OAa \Big).
\ee   
In the second equality we write the identity contribution
explicitly and delete the $a=2^N$ term from the sum, as 
denoted by the prime.

We may now show that all of the basis sets are mutually
unbiased:  Substituting Eq. (\ref{Pexpand}) into  
Eq. (\ref{projection2}) yields
\begin{eqnarray}
  \Tr(\PAa \PBb) &=& \nonumber \\ 
 2^{-N} &+&
 4^{-N} \sum_a^{'} \sum_b^{'} \epsinaa \epsinbb \Tr(\OAa\OBb),
\end{eqnarray}
since terms linear in $\OAa$ have vanishing trace.  It follows 
immediately that if $A$ and $B$ refer to different basis sets, 
then  Eq. (\ref{projection2}) is satisfied.  If $A=B$,
then only the $a=b$ term in the sum survives and
\begin{eqnarray}
   \Tr(\PAa \PAb) &=& 2^{-N} \Big(1 + 
   \sum_a^{'} \epsinaa \epsinba \Big) \nonumber \\ &=& 
  2^{-N} \sum_a \epsinaa \epsinba ~= ~\delta_{\alpha \beta},
\end{eqnarray}
where the orthogonality of $\epsaa/\sqrt{2^N}$ was used. This 
establishes that the $2^N - 1$ basis sets generated (uniquely) 
by the commuting subsets of Pauli operators are in fact 
unbiased. So there is a one-to-one correspondence between 
partitionings of Pauli operators and choices of unbiased 
basis sets. 

\section{Examples for two and three qubits}

We now illustrate this correspondence for systems of two 
and three qubits.  To develop notation, the operator subsets 
for the {\it one} qubit case consist of single elements, 
$\sx$, $\sy$, and $\sz$.  Corresponding basis sets are 
denoted by (x), (y), and (z), where each basis set consists
of the two states ``up'' and ``down'' along the indicated 
axis.  The individual basis states are denoted by $\ket{n_x}$, 
$\ket{n_y}$, and $\ket{n_z}$, where $n_x = 1$ or 0 for spin
``up'' and ``down'' respectively.   The inner products 
between any two states appearing in these basis sets obey
Eqs. (1) and (2).   Obviously, measurements by any of the 
above operators on an eigenstate of any other will produce 
perfectly random results (i.e., an average spin projection 
of zero).  

In the case of two qubits, the dimension of the Hilbert space
is $d = 4$, so that five unbiased basis sets exist.  Fig. 1 
shows these together with the five corresponding operator sets, 
each consisting of three compatible operators.  Subscripts 
indicate three product bases, $(zz)_\pi$, $(xy)_\pi$, and 
$(yx)_\pi$, whose individual states are denoted in the 
(zz)$_{\pi}$ case, for example, by $\kett{n^1_z}{n^2_z}$.  
There are two Bell bases, $(zx)_B$ and $(yz)_{Bi}$, and the 
states belonging to each of these may be written, 
respectively, as 
\be
  \kettp{\naz}{\nbx}{\pm} = \irt \big(\kett{\naz}{\nbx} 
         \pm  \kett{\nnaz}{\nnbx}\big),
\label{Bell1}
\ee
\be
  \kettp{\nay}{\nbz}{\pm i} = \irt \big(\kett{\nay}{\nbz} 
         \pm i \kett{\nnay}{\nnbz}\big),
\label{Bell2}
\ee
where bars denote spin flips; ie, if $n_x = 1$~or 0, then 
$\bar{n}_x = 0$~or 1, respectively. Thus, the four individual 
basis states are explicitly enumerated [in Eq. (\ref{Bell2}), 
for example] by $\kettp{1_y}{1_z}{\pm i}$ and 
$\kettp{1_y}{0_z}{\pm i}$. The factor of $i$ [as denoted by 
the subscript in the basis label $(yz)_{Bi}$]
is not arbitrary; its presence is dictated by the operators 
that define the basis, or equivalently by the requirement 
that the two Bell bases be mutually unbiased.  It is a 
property of Bell bases that they can appear equally simple
if other quantization axes are chosen in appropriate 
combinations.  For example, the Bell bases 4 and 5 in Fig. 1
may be written as $(yy)_B$ and $(xx)_B$, respectively.  While
these are simply different ways of writing the same basis 
sets, truely different alternatives involving all five basis 
sets exist for two qubits (see Refs. \cite{bz2} and 
\cite{Vatan}).  These alternatives also consist of three 
product bases and two Bell bases.

\begin{figure}
\centerline{\psfig{width=6cm,file=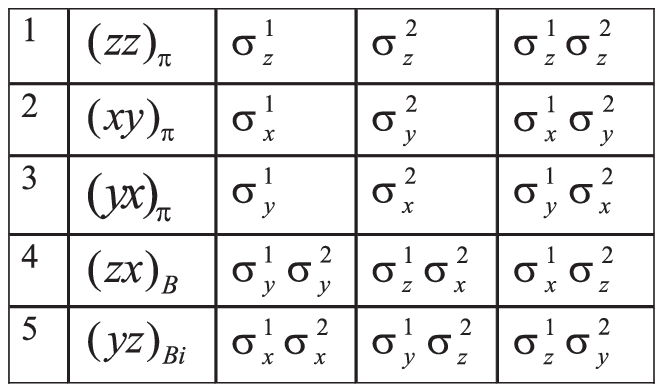}}
\caption{Five unbiased bases sets and corresponding Pauli 
operator sets. Each operator set consists of three commuting 
members, any two of which determine the corresponding basis 
set as their joint eigenbasis.} 
\label{table1}
\end{figure}

The three-qubit Hilbert space has $d = 8$, and thus nine 
unbiased basis sets.  One choice contains three product bases, 
$(xyz)_\pi$, $(yzx)_\pi$, and $(zxy)_\pi$; and six bases 
consisting of maximally entangled states, $(xxx)_{Gi}$,
$(yyy)_G$, $(zzz)_G$, $(xzy)_G$, $(yxz)_G$, and $(zyx)_G$.
The nine basis sets are listed in Fig. 2 and represented
graphically in Fig. 3.   The entangled 
basis sets are labeled by coordinate axes in which the states 
reduce to the familiar Greenberger-Horne-Zeilinger \cite{ghz} 
(GHZ) form.  For example, all of the states belonging to the 
basis set $(zyx)_G$ may be written as
\be
  \ketttp{\naz}{\nby}{\ncx}{\pm} = \irt
  \big(\kettt{\naz}{\nby}{\ncx} \pm
         \kettt{\nnaz}{\nnby}{\nncx}\big);
\label{GHZ}
\ee
these would require more complicated expressions if referred
to other coordinate axes.  Fig. 2 lists the seven-member 
operator sets that correspond uniquely to each basis set. As
in the two-qubit case, the operators involving only a single
Pauli matrix are exhausted within the product basis sets.   

It is striking that in the progression from one to two to
three qubits, the number of totally entangled bases can 
grow from none to two to six, while the number of product
bases remains fixed at three.  It is easy to convince 
onesself that the maximum number of product bases remains
fixed at three for all numbers $N$~of qubits.  

\begin{figure}
\centerline{\psfig{width=9cm,file=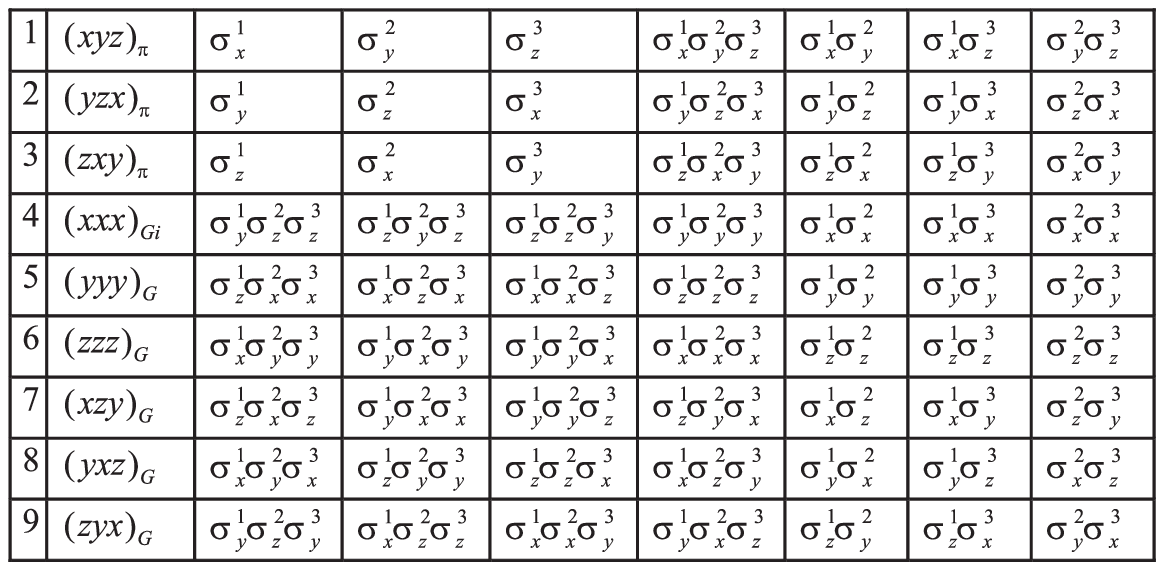}}
\caption{Listing of nine unbiased basis sets and corresponding 
operator sets, each consisting of seven commuting members.  
Particular subsets of three determine the corresponding basis 
sets completely.}
\label{table2}
\end{figure}

\begin{figure}
\centerline{\psfig{width=5cm,file=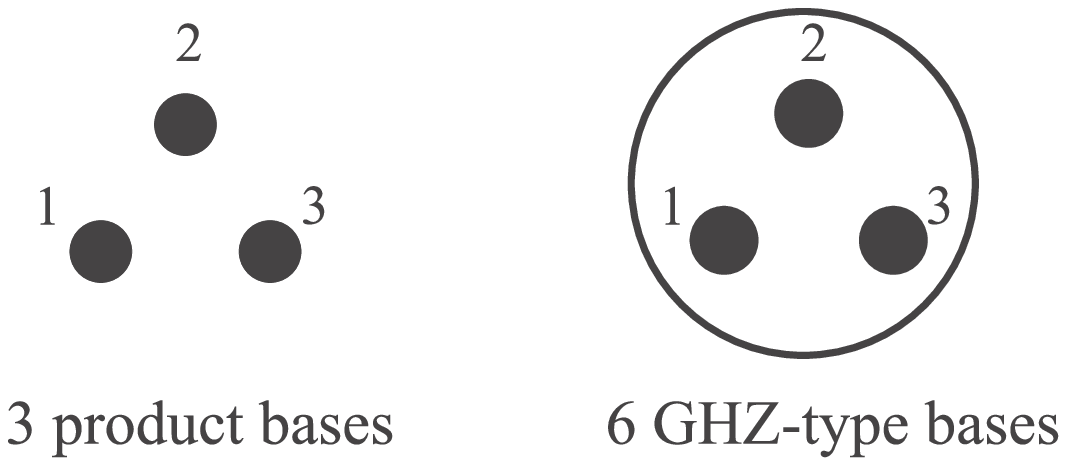}}
\caption{Schematic of unbiased basis sets listed in Fig. 2
- three product and six GHZ bases. The three particles in 
the circle are maximally entangled (in a GHZ state).}
\label{choice1}
\end{figure}

To show that the structure is more flexible with three 
qubits than with two, we describe now a different choice 
of unbiased basis sets for
three qubits, one that cannot be obtained from the previous 
choice by local unitary transformations. In this choice
(Fig. 4), there are no product states, and no states with 
three-particle entanglement.  Every basis consists of states
that are products of one-particle states with Bell states;
one particle is unentangled while the other two are totally
entangled, as depicted in Fig. 5.   The basis sets form 
groupings of three: $(x^1)(yz)_{Bi}$, $(y^1)(zx)_B$, 
$(z^1)(xy)_B$, then $(x^2)(xy)_B$, $(y^2)(yz)_{Bi}$, 
$(z^2)(zx)_B$, and finally $(x^3)(zx)_B$, $(y^3)(xy)_B$,
$(z^3)(yz)_{Bi}$, in which a different particle is
factored out within each group.  Coordinate axes are 
permuted within each group, but not from group to group.  
Factors of $i$~appear once within each group. 

\begin{figure}
\centerline{\psfig{width=9cm,file=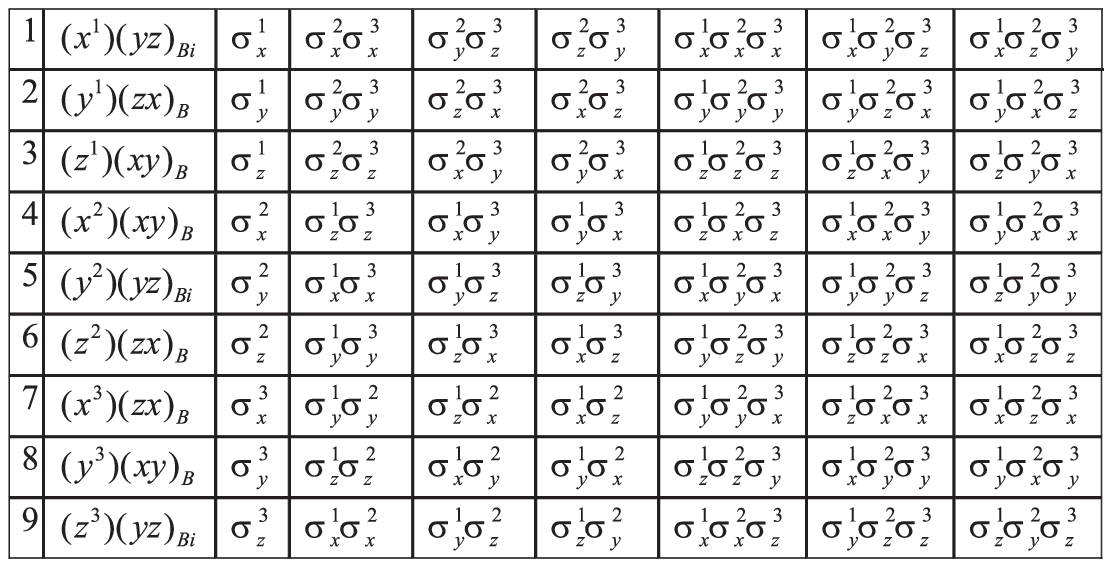}}
\caption{Same structure as in Fig. 3, but all bases are 
partially entangled.  In each of three groups a different 
particle must be singled out as unentangled.} 
\label{table3}
\end{figure}

\begin{figure}
\centerline{\psfig{width=7cm,file=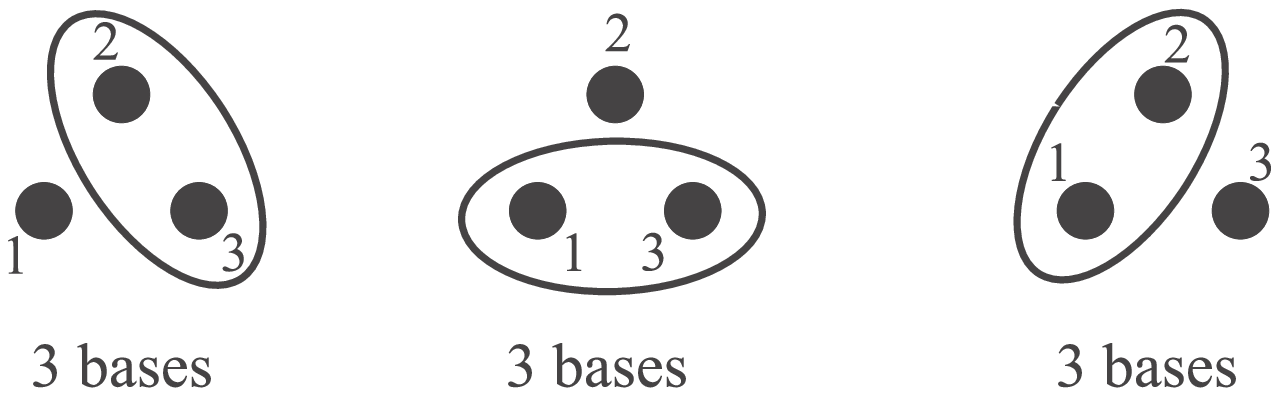}}
\caption{Schematic of unbiased basis sets listed in Fig. 4, 
all partially entangled.  One particle is unentangled with 
the other two, which are totally entangled in Bell states.}
\label{choice2}
\end{figure}

Note that within each grouping, we find {\it three} unbiased
Bell-type bases - a feature that was not seen in the
two-qubit system.  Indeed, if one were to begin with 
three unbiased Bell bases in a two-qubit system, one
could then {\it not} find two additional basis sets.  This
can be understood in terms of the operator decomposition:
The nine operators exhausted by three Bell bases do not 
leave six operators that are decomposable into two commuting 
subsets.  

We also note that the choices of unbiased basis sets given
for the three-qubit case are not obtained from the 
algorithmic construction given in Ref. \cite{wf}.   
Wootters has pointed out \cite{wootters2} that this 
construction  produces another choice which consists of two 
product, four GHZ, and three ``product-Bell'' bases.  In the 
case of two qubits, not surprisingly, the same construction 
produces three product and two Bell bases \cite{wootters2}.

\section{Conclusions}

The two- and three-qubit cases illustrate general points made
at the beginning.  First, with regard to state preparation, 
one can see that $N$ observables suffice to define any of the 
listed basis states completely, representing these states as 
binary numbers.  In the two-qubit case, any two of the
three compatible observables within a subset may be chosen.  
In the three-qubit case, there are many choices of three 
observables that suffice - for example, the first three listed 
within each subset.  In the $N$-qubit case,
we introduced an $\epsaa$ matrix in which an appropriate 
choice of $N$ rows (representing $N$ operators) describe all 
$2^N$ basis states as binary nunbers.

Second, with regard to the determination of a general, possibly 
mixed state, recall that $4^N - 1$ real parameters (15 for two 
qubits and 63 for three qubits) are required to specify the 
$N$-qubit density matrix completely \cite{wootters1,ivanovic}.  
And exactly this number is provided, either by the expectation 
values of the operators themselves, or by all the independent 
probabilities associated with the unbiased basis states.  As we
have proven for the general case, the $4^N-1$ Pauli operators 
can be partitioned into $2^N+1$ subsets, each consisting of 
$2^N-1$ internally commuting observables. The set of all such 
partitionings has a one-to-one correspondence with choices of 
$2^N+1$ unbiased basis sets in the $N$-qubit Hilbert space. 
There are many such choices, and for $N > 2$ the entanglement 
may be distributed over basis sets in many different ways.  The 
maximum number of product bases is fixed at three for any $N$.

The correspondence between basis sets and observables makes it 
possible to regard all Pauli operator subsets within a given 
partitioning as being mutually unbiased:  If the system is 
prepared in a joint eigenstate of one observable set, then it
has a uniform probability distribution over the joint 
eigenstates of any other observable set in the partitioning.  
As a result, all observables outside the original maximal
commuting subset yield minimal information - measurement 
outcomes are uniformly distributed over all possibilities. 

The concept of unbiasedness between observable sets extends the
idea of complementarity of two individual observables that fail 
to commute.  Clearly two such observables must always belong to 
different mutually unbiased subsets within any partitioning.  
However, as the two- and three-qubit examples show, two 
commuting observables may belong to the same or to different 
unbiased subsets.  Their compatibility is dependent upon the 
partitioning. 

{\it Note added.} After this work was completed, an e-print
\cite{Vatan} appeared reporting work which is related to this
work, but complementary in several respects.  Ref. \cite{Vatan} 
obtained a general relationship between complete bases of 
unitary operators (beloning to the general Pauli group) and 
unbiased basis sets, for any power-of-prime dimension.  In this 
paper, we considered the many-qubit case.  We expanded upon 
the physical interpretation of the concept of complementarity.  
We showed that many alternative paritionings are possible and, 
most importantly, entanglement is distributed among unbiased 
basis sets in a partition-dependent manner for $N > 2$.  

\vskip.4truecm
\centerline{{\bf Acknowledgments}}
\vskip.3truecm

We would like to thank W. K. Wootters for insightful and 
informative discussions. \v{C}.B. has been supported by the 
Austrian Science Foundation (FWF), Project No. F1506, and by 
the QIPC Program of the European Union.  J.L. thanks the Erwin 
Schr\"{o}dinger Institute for its hospitality during the period 
when this work was carried out.

\end{document}